\documentclass[11pt]{article}
\usepackage[T1]{fontenc}
\usepackage[superscript]{cite}
\usepackage{authblk}
\usepackage[usenames,dvipsnames,svgnames,table]{xcolor}
\usepackage{outlines}
\usepackage{amsmath} 
\usepackage{amssymb} 
\usepackage{mathtools}
\usepackage[nottoc]{tocbibind}
\usepackage{enumitem}
\usepackage{graphicx} 
\usepackage[margin=0.9in]{geometry}
\usepackage{lineno}
\usepackage{times}
\usepackage{flafter}
\edef\restoreparindent{\parindent=\the\parindent\relax}
\usepackage{parskip}
\restoreparindent
\usepackage{hyperref}
\hypersetup{
    hidelinks,
    colorlinks = true,
    bookmarksnumbered,
    citecolor = Blue,
    linkcolor = Blue,
    urlcolor  = Blue,
    hypertexnames=false
}
\usepackage[capitalise,nameinlink]{cleveref}
\creflabelformat{equation}{#2#1#3}
\usepackage{autonum}
\usepackage[labelfont=bf,font=small,margin=0.5cm]{caption}
\usepackage{nameref}
\usepackage{titlesec}

\setenumerate[1]{label=\Alph*.} 
\setenumerate[2]{label=\Roman*.}
\setenumerate[3]{label=\alph*.}
\setenumerate[4]{label=\roman*.}

\newcommand{\scale}[0]{\sigma}
\newcommand{\cutoff}[0]{\mu}
\newcommand{\immbi}[0]{r}

\title{\huge
Planned behavior, perceptual biases, \\and the dynamics of collective action
}
\author[1,*]{Alice C.~Schwarze}
\author[2,3,*]{Mari Kawakatsu}
\author[4]{Sarah Iams}
\author[5,6,**]{Nina H.~Fefferman}
\author[7]{Tahra L.~Eissa}
\affil[1]{\small Department of Mathematics, Dartmouth College, 
Hanover, NH 03755}
\affil[2]{\small Department of Biology, University of Pennsylvania, 
Philadelphia, PA 19104}
\affil[3]{\small Center for Mathematical Biology, University of Pennsylvania, 
Philadelphia, PA 19104}
\affil[4]{\small John A. Paulson School of Engineering and Applied Sciences, Harvard University, 
Cambridge, MA 02138}
\affil[5]{\small Department of Ecology and Evolutionary Biology, University of Tennessee, 
Knoxville, TN 37996}
\affil[6]{\small Department of Mathematics, University of Tennessee, 
Knoxville, TN 37996}
\affil[7]{\small Department of Applied Mathematics, University of Colorado Boulder, 
Boulder, CO 80309}
\affil[*]{\small These authors contributed equally}
\affil[**]{\small Author to whom correspondence should be addressed: nina.h.fefferman@gmail.com, she/her/hers}
\date{\today}

\begin{document}

\maketitle

\begin{abstract}

Many classical models of collective behavior assume that emergent dynamics result from external and observable interactions among individuals. However, how collective dynamics in human populations depend on the internal psychological processes of individuals remains underexplored. Here, we develop a mathematical model to investigate the effects of internal psychology on the dynamics of collective action. Our model is grounded in the theory of planned behavior---a well-established conceptual framework in social psychology that links intrinsic beliefs to behavior. By incorporating temporal biases in social perception and individual differences in decision-making processes into our model, we find that the interplay between internal and external drivers of behavior can produce diverse outcomes, ranging from partial participation in collective action to rapid or delayed cascades of action-taking. 
These distinct outcomes are preceded by transient dynamics that are qualitatively similar to one another, which, just as in real-world scenarios, makes it difficult to predict long-term collective dynamics from early observations. Our model thus provides a useful test bed for methods that aim to predict the emergence of collective action, and it lays the groundwork for studying the nuanced dynamics of collective human behavior arising from the interaction between psychological processes and observable actions.

\end{abstract}

\newpage
\section{\centering I. INTRODUCTION} 

From pandemic management to climate change, societal challenges often require individuals to act collectively.
But understanding when and how the actions of a few give rise to collective action is challenging, in part because of the intricate interplay between internal and external drivers of social behavior.  
For example, the decision to vote may be driven by an intrinsic belief in civic duty, but it may also be influenced by observing civic-minded neighbors displaying “I voted” stickers \cite{bond2017social}. Similarly, the decision to be vaccinated may be driven by a personal concern about the risks of a pathogen, but it may also be influenced by discussions with peers who fear manipulation by health authorities \cite{jennings2021lack}. 
While these decisions involve a binary choice between \textit{whether} or not to act, other decisions may involve more than two options. For instance, in stock markets, traders observe each other's value propositions about \textit{when} and \textit{how} to act; this produces a communally constructed understanding of the 
value of a stock, which in turn influences the traders' decisions to buy or sell that stock  (see \cite{cont2000herd} and references therein). Understanding such bidirectional interplay between internal, psychological processes and external, observable behaviors is a fundamental challenge in applying complex systems research to human behavior\cite{rollin2021s,beckage2018linking,silk2022observations}. 

Classical models of complex systems typically assume that collective dynamics arise from observable interactions among individuals. This assumption is appropriate for phenomena driven by reactions to the physical movements of nearby individuals \cite{katz2011inferring}, such as flocking in birds \cite{bialek2012statistical} or schooling in fish \cite{tunstrom2013collective}. Many models of emergent dynamics in human groups---including simple contagion \cite{wang2016computational}, complex contagion \cite{iacopini2019simplicial}, social influence \cite{schwenk2008simple}, and persuasion \cite{wang2017heuristic}---also focus on external drivers of behavior, making the assumption that individuals' decisions to act depend on local exposures to the actions taken by their neighbors.
While useful, these models do not fully capture how social influence can shape behavior. In particular, there is a need to complement observable interactions between individuals with psychological processes within individuals that govern (1) how the actions of others are perceived, (2) how those perceptions shape an individual's intention to act, and (3) whether or not that intention results in an action. 

The \textit{theory of planned behavior} from social psychology offers a conceptual framework that links these psychological dynamics to behavior.
This well-established framework posits that an individual's intention to take an action is shaped by three key components: internal attitude, defined as the individual's intrinsic belief about whether to take the action or not; subjective social norm, defined as the behavioral norm perceived from the observed behavior of others; and perceived behavioral control, defined as the perceived degree of control over an intended behavior (or its outcome) \cite{ajzen1991theory, ajzen1985intentions}. 
This theory provides a basis for studying how group outcomes depend on factors that shape intrinsic dynamics, including perceptual biases and individual differences in the integration of external information. 
While recent studies on belief formation have examined the role of biases (e.g., temporal biases\cite{Feinstein2021}\! ; bounded confidence\cite{Lorenz2007,PorterGleeson2016,ChuLiPorter2024,Koertje2024}, i.e., biases toward like-minded individuals) and individual variation (e.g., behavioral diversity in networks where opinions and links co-evolve\cite{Sayama2020,Bullock2023}), their impact on action-taking dynamics remains underexplored.

Here we draw on the theory of planned behavior to study how intrinsic dynamics affect the frequency and timing of action-taking in social groups. 
We develop a mathematical model of collective action-taking that incorporates the three components of the theory of planned behavior---internal attitude, subjective social norm, and perceived behavioral control. To examine how temporally biased perceptions affect collective dynamics, we also introduce an immediacy bias---a bias toward the most recent observations \cite{murdock1962,tetlock1983accountability}. We show that temporal bias in observations and heterogeneity in internal processes lead to diverse collective outcomes, including partial or delayed cascades of action-taking at the population level and modulations in the frequency of action-taking at the individual level. Our model thus produces richer and more realistic dynamics than traditional models that assume that focus on external drivers of behavior and neglect the effects of internal psychology.
Our approach lays the groundwork for studying the nuanced dynamics of collective human behavior arising from the interplay between simple action rules and perceptual biases.\footnote{Our code repository for this paper is available under \href{https://github.com/acuschwarze/temporally-biased-planned-behavior}{https://github.com/acuschwarze/temporally-biased-planned-behavior}.}

\section{\centering II. MODEL} \label{sec:model}

Drawing on the theory of planned behavior, we develop a piece-wise continuous, deterministic model of collective action-taking behavior. Our model incorporates variables that capture individuals' internal attitudes, perceived social norms, and perceived behavioral control, as well as parameters that capture the effects of immediacy (i.e., the most recent observations) on behavior. We will first introduce a general model with $M$ possible actions. We will subsequently focus on the special case of $M=1$ to demonstrate the rich dynamics that can arise even from this simple implementation of our model.

\renewcommand{\figurename}{FIG.}
\renewcommand*{\figureautorefname}{FIG.}

\begin{figure}[b!]
    \centering
    \includegraphics[width=0.65\textwidth]{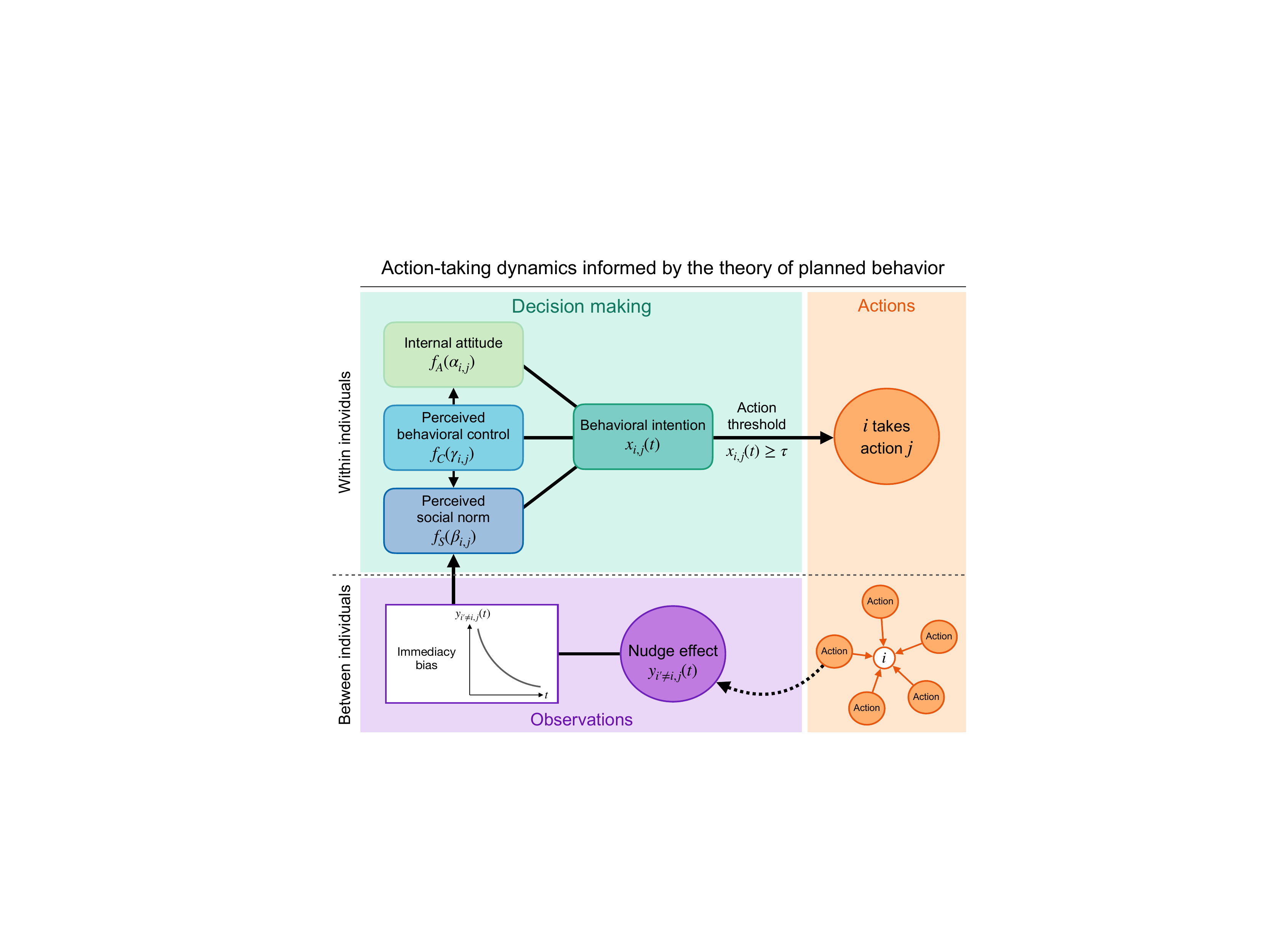}
    \caption{
    \textbf{A model of collective action-taking informed by the theory of planned behavior.}
    The $M$-action model is comprised of population observations that modulate an individual's internal perceptions of social norms, which, when combined with internal attitude and perceived behavioral control, comprise the individual's behavioral intention. When this intention reaches a threshold, an individual takes an action, which can then be observed by other members of the population. 
    }
    \label{fig:schematic}
\end{figure}

\subsection{\centering A. The $M$-action model}

We consider a population of $N$ individuals, each of whom can take one of $M$ actions at a given time $t$. An individual is characterized by their intention to act: 
\textit{behavioral intention} $x_{i,j}(t)\in(-1,1)$ captures how strongly individual $i$ intends to take action $j$ at time $t$. 
Individuals take actions whenever their behavioral intentions are sufficiently strong, as described below. Taking actions influences others' intentions to act: the \textit{nudge effect} $y_{i,j}(t)\in[0,1]$ captures how strongly individual $i$'s actions up to time $t$ nudge other individuals to take action $j$. The population is thus described by $2NM$ dynamical variables, $x_{i,j}$ and $y_{i,j}$, with $i\in\{1,\dots,N\}$ and $j\in\{1,\dots,M\}$.

To track the number of actions taken over time, we also define \textit{action count} $z_{ij}(t)\in\mathbb Z_+$ as the number of times individual $i$ has taken action $j$ up to time $t$. We initialize the action count as zero for all actions, i.e., $z_{i,j}(0)=0$ for all individuals $i$ and actions $j$. 

\subsubsection*{\centering \textbf{\textit{1. Dynamics of behavioral intentions and nudge effects}}}
The theory of planned behavior postulates that behavioral intentions are influenced by three interacting types of behavioral perceptions within an individual: their internal attitude, their perceived social norm, and their perceived behavioral control \cite{ajzen1991theory} (\autoref{fig:schematic}). Based on this, we model the dynamics of the behavioral intentions $x_{i,j}(t)$ using the following system of ODEs:
\begin{equation+} 
    \dot x_{i,j} = \left[f_A\left(\alpha_{i,j}\right) + f_S\left(\beta_{i,j}\right)\right] f_C\left(\gamma_{i}\right) g(x_{i,j})\;, \qquad i,j\in\{1,\dots,N\} \;, \label{eq:mmodel-xdot}  
\end{equation+} 
where the functions $f_A$, $f_S$, and $f_C$, described below, capture the effects of internal attitudes, perceived social norms, and perceived behavioral control, respectively. We implement $f_C(\gamma_i)$ as a multiplicative factor because while perceived behavioral control can amplify the effects of intrinsic motivations (internal attitude) or social drive (perceived social norm), it alone cannot boost behavioral intentions (i.e., if $f_A(\alpha_{i,j}) + f_S(\beta_{i,j})=0$, then $\dot x_{i,j}=0$ regardless of the value of $f_C(\gamma_i)$). The function $g(x_{i,j})$ governs the intrinsic logistic growth of behavioral intentions:
\begin{equation+}
    g(x_{i,j}) = (1-x_{i,j})(1+x_{i,j}) \:.
\end{equation+} 

\subsubsection*{\centering \textbf{\textit{2. Internal attitudes}}}

The \textit{internal attitude} parameter $\alpha_{i,j}\in(-1,1)$ captures individual $i$'s intrinsic propensity toward taking action $j$ (\autoref{fig:schematic}). A larger value of $\alpha_{i,j}$ means that $i$ will take action $j$ more readily even in the absence of social influence. The effect of internal attitudes $\alpha_{i,j}$ on behavioral intentions $x_{i,j}$ is determined by the function $f_A$. According to the theory of planned behavior, internal attitudes drive self-reinforcement in behavioral intentions \cite{ajzen1985intentions}. To reflect this, we assume that $f_A$ satisfies $f_A(0)=0$ and $f_A'(\alpha_{i,j})> 0$ for all $\alpha_{i,j}\in(-1,1)$; that is, in the absence of social influence (perceived control or perceived social norm), individual $i$'s intention $x_{i,j}$ will grow if $i$'s intrinsic propensity to take action $j$ is positive ($\alpha_{i,j}>0$), and $x_{i,j}$ will decay if $\alpha_{i,j}<0$. 

\subsubsection*{\centering \textbf{\textit{3. Perceived social norm}}}

Nudge effects shape the social norms that individuals perceive toward an action, which, in turn, affects their intentions to act (\autoref{fig:schematic}). To measure the strength of the perceived social norm associated with specific actions, we define the time-dependent \textit{nudge coefficient} $\beta_{i,j}\in[0,1]$ as the average nudge effect toward action $j$ experienced by an individual $i$:
\begin{equation+}
    \beta_{i,j}=\frac{1}{N-1}\sum_{\substack{i'=1,\\i'\neq i}}^N y_{i',j}(t)\:.
\end{equation+}
The effect of the nudge coefficient on behavioral intentions is captured by the function $f_S\left(\beta_{i,j}\right)$, which we assume is increasing in $\beta_{i,j}$; that is, the larger the nudge coefficient $\beta_{i,j}$, the stronger the social expectations to take action $j$, and thus the faster the growth of intention $x_{i,j}$.

\subsubsection*{\centering \textbf{\textit{4. Perceived behavioral control}}}

Finally, the \textit{perceived-control coefficient} $\gamma_{i}\in[0,1]$ measures how easy an individual $i$ believes it will be to take actions. We operationalize this measure as the average level of activity observed by a focal individual among all others in the population:
\begin{equation+}
    \gamma_{i}=\frac{1}{(N-1)M}\sum_{\substack{i'=1,\\i'\neq i}}^N\sum_{\substack{j'=1}}^M y_{i',j'}\:.
\end{equation+}
In other words, the strength of behavioral control $\gamma_i$ that a focal individual $i$ infers is the observed nudge effects averaged over all other individuals and over all possible actions. 

\subsubsection*{\centering \textbf{\textit{5. Action threshold and immediacy}}}

Actions occur whenever the corresponding behavioral intentions reach or exceed a prescribed threshold: individual $i$ takes action $j$ if $x_{i,j}\geq\tau$, where $\tau$ is the \textit{action threshold} $\tau\in\mathbb{R}$. When $i$ takes action $j$ at time $t$, the variables $x_{i,j}$, $y_{i,j}$, and $z_{i,j}$ undergo a discontinuous change, given by
\begin{align+}  
    x_{i,j}(t+dt) & = 0 \nonumber \;, 
    \\
    y_{i,j}(t+dt) & = 1\;, 
    \\
    z_{i,j}(t+dt) & = z_{i,j}(t) + 1 \nonumber  \;.
\end{align+}

Immediacy bias describes the cognitive tendency to give more weight to information encountered most recently \cite{murdock1962}. In the context of our model, this means that nudge effects $y_{i,j}$ decay over time; for simplicity, we assume that this decay follows an exponential function, given by
\begin{equation+}
    \dot y_{i,j} = -\immbi y_{i,j}\,,
\end{equation+} 
where $\immbi\geq0$ is the \textit{immediacy} parameter. When immediacy bias is absent (i.e., $\immbi=0$), nudge effects remain constant between actions; when immediacy bias is strong (i.e., large $\immbi$), nudge effects decay quickly. 

\renewcommand{\thetable}{\Roman{table}}

\begin{table}[h!]
    \centering\small
    \renewcommand{\arraystretch}{1.10}
    \begin{tabular}{cll} 
    \hline
        \textbf{Symbol} & \textbf{Name} & \textbf{Constraint} \\
    \hline
        $N$ & Population size & $N\in\mathbb Z_+$\\
        $M$ & Number of actions & $M\in\mathbb Z_+$\\
        $x_{i,j}$ & Behavioral intention & $x_{i,j}\in(-1,1)$ \\
        $y_{i,j}$ & Nudge effect & $y_{i,j}\in[0,1]$ \\
        $z_{i,j}$ & Action count & $z_{i,j}\geq0$ 
        \\
        $\alpha_{i,j}$ & Internal attitude & $\alpha_{i,j}\in(-1,1)$ \\
        $\beta_{i,j}$ & Nudge coefficient & $\beta_{i,j}\in[0,1]$ \\
        $\gamma_i$ & Perceived-control coefficient & $\gamma_i\in[0,1]$ \\
        $\sigma_A$ & Strength of internal attitudes & $\sigma_A\geq 0$ \\
        $\sigma_S$ & Strength of social norms & $\sigma_S\geq 0$ \\
        $\sigma_C$ & Strength of perceived behavioral control & $\sigma_C\geq 0$ \\
        $\immbi$ & Immediacy parameter & $\immbi\geq 0$ \\
        $\mu_S$ & Social norm threshold & $\mu_S\in[0,1]$ \\
        $\mu_C$ &  Baseline perceived control & $\mu_C\geq 0$\\
    \hline
    \end{tabular}
    \caption{Key parameters and variables in the model.}
    \label{tab:params}
\end{table}

\subsection{\centering B. A $1$-action model}

While our model can accommodate any number of individuals and actions and a variety of functional forms for the dynamics of behavioral intentions, it can produce rich dynamics even in a simple scenario of interest, namely with a single action and linear functions $f_A$, $f_S$, and $f_C$. 

To demonstrate this, we implement a 1-action ($M=1$) model, described by the $2N$ dynamical variables $x_i(t)$ and $y_i(t)$ for $i\in\{1\,\dots, N\}$. 
We choose the following functions:
\begin{align+}
        f_A(\alpha_i)& =\scale_A\alpha_i \;, \nonumber \\
        f_S(\gamma_i)& =\scale_S(\gamma_i-\cutoff_S) \;, \\
        f_C(\gamma_i)& =\scale_C(\gamma_i+\cutoff_C) \;. \nonumber
\end{align+}
The parameters $\scale_A$, $\scale_S$, and $\scale_C$ tune the strength of the effects of internal attitudes, perceived social norms, and perceived behavioral control on behavioral intentions. The parameter $\cutoff_S$ indicates the level of collective action-taking that leads to positive perceived social norms. For example, when $\immbi=0$, the value of $\cutoff_S$ indicates the fraction of individuals that need to take an action before perceived social norms increase individuals' intentions to take an action. The parameter $\cutoff_C$ indicates the base level of perceived behavioral control when no one has taken an action. When $\cutoff_C=0$, a stable fixed point of the population is $x_i=y_i=0$ for all individuals $i$; in other words, unless the dynamical system starts from an initial state that is sufficiently far from that trivial fixed point, all individuals in the population remain inactive for all time.

When $M=1$, individuals have a binary choice between taking an action or remaining inactive at every time $t$, and each nudge coefficient $\beta_{i,j}$ is equal to the corresponding control coefficient $\gamma_i$. 
Thus, the dynamics of $x_i$ and $y_i$ are governed by
\begin{align+}
    \dot x_{i} & = \left[\scale_A\alpha_{i}  + \scale_S\left(\gamma_{i}-\cutoff_S\right) \right] \scale_C\left(\gamma_{i}+\cutoff_C\right) \left(1-x_{i}\right)\left(1+x_{i}\right) \;, \label{eq:1model-xdot}
    \\
    \dot y_{i} & = -\immbi y_{i} \;, \label{eq:1model-ydot}
\end{align+}
with discontinuous updates
\begin{align+}    
    x_{i^*}(t+dt) & = 0 \;,  \nonumber
    \\
    y_{i^*}(t+dt) & = 1\;,
    \\
    z_{i^*}(t+dt) & = z_{i^*}(t) + 1 \;. \nonumber
\end{align+}
whenever $i^*$ takes action (i.e., $x_{i^*}(t)\geq\tau$).

\section{\centering III. RESULTS} \label{sec:results}

\subsection{\centering A. The emergence and dynamics of action-taking cascades}

Our overarching goal is to understand how perceptual nuances embedded in the model---social norm, behavioral control, and immediacy bias---affect the emergence and dynamics of action-taking cascades.
To study these two aspects of behavioral cascades, we must first understand which individuals, if any, in a given population take actions. We therefore begin by asking two questions: 
How does variation among the intrinsic action-taking propensities in a population, as captured by internal attitudes $\alpha_i$, determine who takes actions in the long run? And how frequently will individuals act?

\begin{figure}[b!]
    \centering
    \includegraphics[trim=1cm 1.2cm 1cm 1.2cm, width=0.67\textwidth, clip]{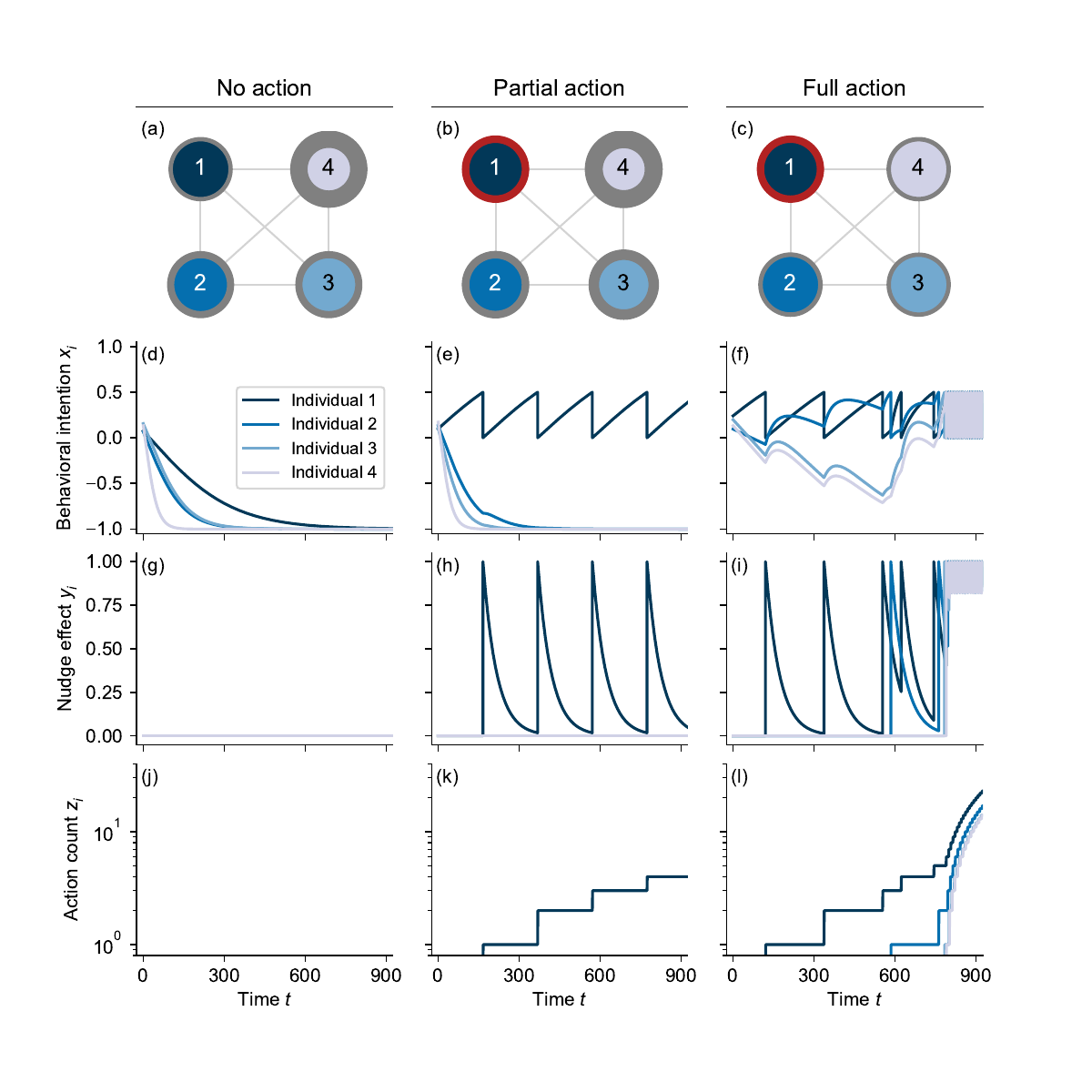}
    \caption{{\bf The number of action-taking individuals depends on their internal attitudes.} 
    We performed numerical simulations of the $1$-action model with $N=4$ individuals and with different distributions of internal attitudes $\alpha_i$.
    Columns show three qualitatively distinct outcomes: `no action' (left column), `partial action' (middle column), and `full action' (right column).
    (a)--(c): 
    Circles with red (gray) edges indicate individuals whose internal attitude is larger than (smaller than) the social norm based on the baseline threshold $\sigma_S\cutoff_S/\scale_A$ of an inactive population. Edge widths denote the magnitude of the difference between their internal attitude and baseline threshold $|\alpha_i-\sigma_S\cutoff_S/\scale_A|$. (d)--(l): Behavioral intentions $x_i$ (d--f), nudge effects $y_i$ (g--i), and action counts $z_i$  (j--l) of individuals $i\in\{1,2,3,4\}$, indicated by the colors. 
    Parameters: $\scale_A=1$, $\scale_S=1$, $\scale_C=0.05$, $\cutoff_S=\cutoff_C=0.5$, $\immbi=0.02$. 
    Initial conditions: $x_i(0)\sim U(0, 0.25)$ and $y_i(0)=0$.}
    \label{fig:scenarios}
\end{figure}

To demonstrate how variation in internal attitudes impacts long-term action taking, we numerically simulate the 1-action model with $N=4$ individuals with different sets of internal attitudes $\alpha_i$.
We find three qualitatively distinct collective outcomes: the `no action' scenario, in which no one takes action (\autoref{fig:scenarios}a); the `partial action' scenario, in which a subset of individuals take action (\autoref{fig:scenarios}b); and the `full action' scenario, in which all individuals take action (\autoref{fig:scenarios}c).

In the `no action' scenario (\autoref{fig:scenarios}a), individuals 1--3 have positive internal attitudes $\alpha_i$, but all values of $\alpha_i$ fall short of the social-norm baseline threshold $\sigma_S\cutoff_S/\scale_A$ that would trigger an individual to increase their intention $x_i$. As a result, the intentions of all individuals decrease with time (\autoref{fig:scenarios}d), and the population remains inactive (\autoref{fig:scenarios}g,j).

In the `partial action' scenario (\autoref{fig:scenarios}b), individual 1 has an internal attitude that surpasses the baseline social-norm threshold ($\alpha_1>\sigma_S\cutoff_S/\scale_A$) such that their intention increases from time $t=0$ and leads to individual 1 taking actions (\autoref{fig:scenarios}e). Individual 1's actions produce positive nudge effects $y_1(t)$ (\autoref{fig:scenarios}h) and a corresponding increase in action count (\autoref{fig:scenarios}k). However, this nudge effect is not sufficient to produce positive intentions amongst the other individuals because their internal attitudes are too small as compared to the social norm threshold, causing their intentions to decrease rapidly. Thus, the population splits into a single active individual (individual 1) and a cluster of inactive individuals (individuals 2--4).

In contrast, in the `full action' scenario (\autoref{fig:scenarios}c), individuals 2--4 have sufficiently large internal attitudes $\alpha_i$ so that, although their intentions initially decrease, the successive nudge effects eventually lead all individuals to take actions (\autoref{fig:scenarios}f). As each additional individual becomes active, nudge effects increase (\autoref{fig:scenarios}i) and all active individuals increase their action-taking frequency (\autoref{fig:scenarios}l). Thus, the emergence of action-taking cascades depends heavily on both the value of each individual's internal attitude and relative recruitment via nudge effects, whereby recruitment of more individuals leads to increased nudge effects and correspondingly, increased action-taking frequency across the population.

\begin{figure}[h!]
    \centering
    \includegraphics[width=0.75\textwidth]{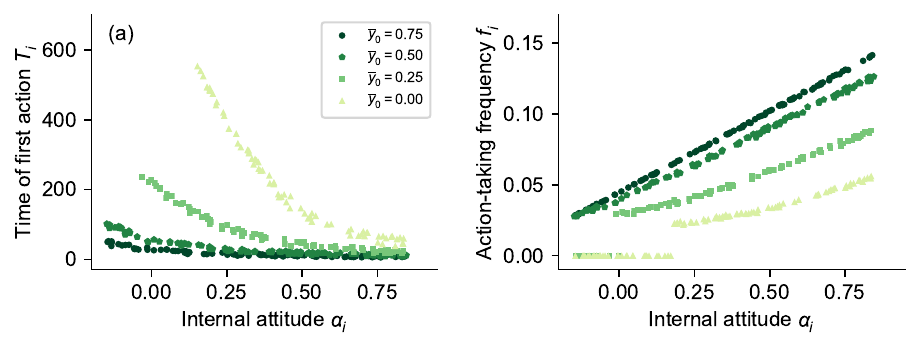}
    \caption{
    {\bf Stronger internal attitudes promote more frequent action-taking.} We performed numerical simulations of the 1-action model with $N=100$ individuals with different initial nudge effects. (a):
    The time-averaged frequencies $f_i$ of actions taken by individuals $i$ as a function of their internal attitudes $\alpha_i$. (b): The times $T_i$ of the first action taken by individuals $i$ as a function of their internal attitudes $\alpha_i$.
    Colors denote mean initial nudge effects $\overline{y}_0$. 
    Parameters: $\scale_A=1$, $\scale_S=1$, $\scale_C=0.05$, $\cutoff_S=0.5$, $\cutoff_C=0.5$, $\immbi=0.02$,  
    $\alpha_i\sim \overline{\alpha}+U(-0.5, 0.5)$. Initial conditions: $x_i(0)\sim U(0, 0.25)$ and $y_i(0)=0$ when $\overline{y}_0=0$ and $y_i(0)\sim\overline{y}_0+U(-0.25, 0.25)$ when $\overline{y}_0>0$.
    }
    \label{fig:attitudes}
\end{figure}

We also investigate the effect of internal attitudes on how frequently individuals act. To do so, we simulate a population of $N=100$ individuals whose internal attitudes $\alpha_i$ are sampled from a uniform distribution. We find that internal attitudes $\alpha_i$ are strongly correlated with which individuals become active, when they take their first action (\autoref{fig:attitudes}a), and how frequently they act (\autoref{fig:attitudes}b). Individuals with large values of $\alpha_i$ are more likely to become active, become active sooner, and tend to act more frequently than individuals with small $\alpha_i$. However, these effects are modulated by the initial nudge effects $\overline{y}_0$. When there are no nudge effects at time $t=0$ ($\overline{y}_0=0$; light green triangles in \autoref{fig:attitudes}), individuals with $\alpha_i< 0.19$ remain inactive throughout, whereas individuals with $\alpha_i\geq 0.19$ become active. However, increasing the initial nudge effects increases the proportion of individuals who become active: individuals with lower internal attitudes become recruited as the initial nudge effect increases ($\overline{y}_0=0.25$ (light green squares), $\overline{y}_0=0.50$ (dark green pentagons), and $\overline{y}_0=0.75$ (dark green circles) in \autoref{fig:attitudes}). Overall, individuals with stronger internal attitudes (larger $\alpha_i$) are more likely to become active and, when they do, act more frequently.

\begin{figure}[t!]
    \centering
    \includegraphics[trim=0cm 0cm 0cm 0cm, width=0.8\textwidth]{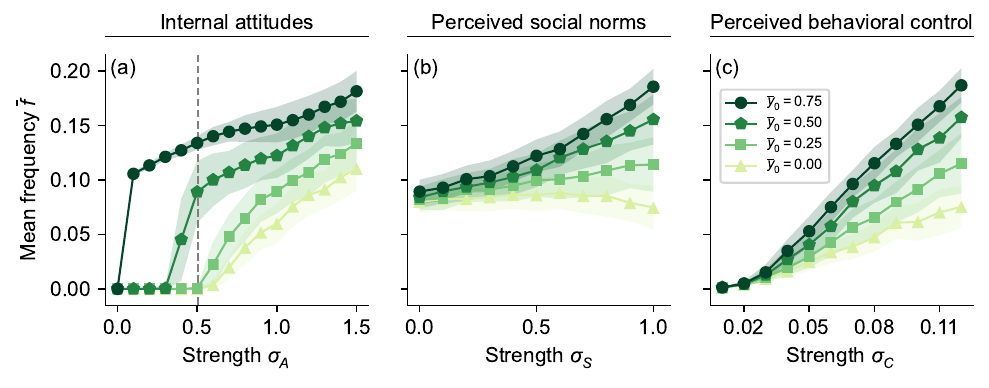}
   \caption{{\bf Increasing the strengths of internal attitudes, perceived social norms, and perceived behavioral control generally increases the mean action-taking frequency.}
   Panels show ensemble mean action-taking frequencies $\overline{f}$ ($\pm$SD) across 100 replicate simulations, each in a population of $N=100$ individuals and with $t=5000$ time steps. 
   Colors denote mean initial nudge effects $\overline{y}_0$. 
   (a) We vary the strength $\scale_A$ of internal attitudes, with $\scale_S=1$ and $\scale_C=0.1$. The gray dashed line shows the critical $\sigma_A$ at which a phase transition occurs for $y_{i}(0)=0$.
   (b) We vary the strength $\scale_S$ of perceived social norms, with $\scale_A=1$ and $\scale_C=0.12$.
   (c) We vary the strength $\scale_C$ of perceived behavioral control, with $\scale_A=1$ and $\scale_S=1$. 
   Other parameters:  
   $\cutoff_S=0.5$, $\cutoff_C=0.5$, $\immbi=0.02$, 
   $\alpha_i\sim U(-0.4, 1)$. Initial conditions: $x_i(0)\sim U(0, 0.5)$ and $y_i(0)=0$ when $\overline{y}_0=0$ and $y_i(0)\sim\overline{y}_0+U(-0.25, 0.25)$ when $\overline{y}_0>0$.}
   \label{fig:asc}
\end{figure}

\subsection{\centering B. Competition between internal attitudes and perceived social norms}

We have shown that internal attitudes are a key driver of action-taking at the individual level. 
To characterize the dynamics at the population level, 
we now consider all three elements of the theory of planned behavior---internal attitudes, perceived social norms, and perceived behavioral control---and study how they impact the collective dynamics of action-taking. To do so, we perform numerical simulations in populations of $N=100$ individuals with varied strengths of the different components, and we compute the mean frequency of action-taking in the population (\autoref{fig:asc}).

In general, the population acts more frequently as the strength $\scale_A$ of internal attitudes increases (\autoref{fig:asc}a).  When the mean initial nudge effects are low (e.g., $\overline{y}_0=0$, $0.25$, or $0.5$), the shift from infrequent to frequent action-taking is marked by a phase transition in $\scale_A$. For subcritical values of $\scale_A$, no individuals take actions (i.e., $\overline{f}=0$). However, as $\scale_A$ crosses a critical value $\scale_A^*$, the mean action-taking frequency increases rapidly with the strength of internal attitudes.  In the absence of initial nudge effects (i.e., $y_i(0)=0$ for all individuals $i$; specifically, $\overline{y}_0=0$ in \autoref{fig:asc}a), the critical value $\scale_A^*$ has a lower bound $\chi_A=\scale_S\cutoff_S/\max_i\left\{\alpha_i\right\}$ (dashed gray line in \autoref{fig:asc}a), which we derive in Appendix A. No actions are taken when $\scale_A<\chi_A$ because no individual in the population has a sufficiently positive internal attitude to take the first action.

Unlike in the case of internal attitudes, we observe no phase transitions associated with the strength $\scale_S$ of perceived social norms. Nevertheless, the mean frequency of action-taking follows qualitatively distinct trends depending on initial nudge effects (\autoref{fig:asc}b). When the mean initial nudge effect $\overline{y}_0$ is zero, the mean frequency of actions $\overline{f}$ decreases monotonically as the strength of perceived norms $\scale_S$ increases (light green triangles in \autoref{fig:asc}b). The absence of nudge effects at time $t=0$ leads to an initial period of inaction during which there is a strong social norm against taking actions, thereby delaying or inhibiting actions. Large $\scale_S$ exacerbates this effect. In contrast, when the initial nudge effect is sufficiently large, the mean frequency increases monotonically with increasing $\scale_S$ (e.g., $\overline{y}_0\geq0.25$ in \autoref{fig:asc}b). This increase is more pronounced for larger initial nudge effects (e.g., $\overline{y}_0=0.75$), as individuals begin taking actions earlier. These earlier actions can nudge individuals with negative internal attitudes to act before their internal attitudes drive their behavioral intentions toward inaction (i.e., $x_i$'s approach $-1$). These effects are enhanced as the strength of perceived social norms increases. 

Finally, we consider perceived behavioral control. The strength $\scale_C$ of perceived behavioral control acts as a time-scale parameter in the dynamics of the intentions $x_i$ (\cref{eq:1model-xdot}). Thus, increasing $\scale_C$ results in a higher average frequency of taking action (\autoref{fig:asc}c). This effect is more pronounced for stronger initial nudge effects (e.g., $\overline{y}_0=0.75$) because individuals begin taking actions early while the intentions $x_i(t)$ change slowly. 

Overall, our results show that when individuals' internal motivations are strengthened, either through internal attitudes or through perceived behavioral control, they tend to act more frequently, regardless of the initial nudge effects. In contrast, the qualitative impact of social influence on action-taking frequency is modulated by the strength of initial nudge effects. In short, while the initial nudge effects modulate the transient dynamics, we find that the initial behavioral intentions have little impact on the long-term behavior of the collective.

\subsection{\centering C. Delay and inhibition of action cascades via immediacy bias}

We have characterized how each component of the theory of planned behavior---internal attitudes, perceived social norms, and perceived behavioral control---affects the frequency of action-taking, assuming that immediacy bias is weak ($r=0.02$) and therefore nudge effects decay slowly. 
However, the strength of immediacy bias may impact whether or not individual actions will induce sustained collective action. To explore this possibility, we analyze the effect of the immediacy parameter $\immbi$ on the emergence of action-taking cascades. 

\begin{figure}[h!]
    \centering
    \includegraphics[trim=0cm 0cm 0cm 0cm, width=0.7\textwidth, clip]{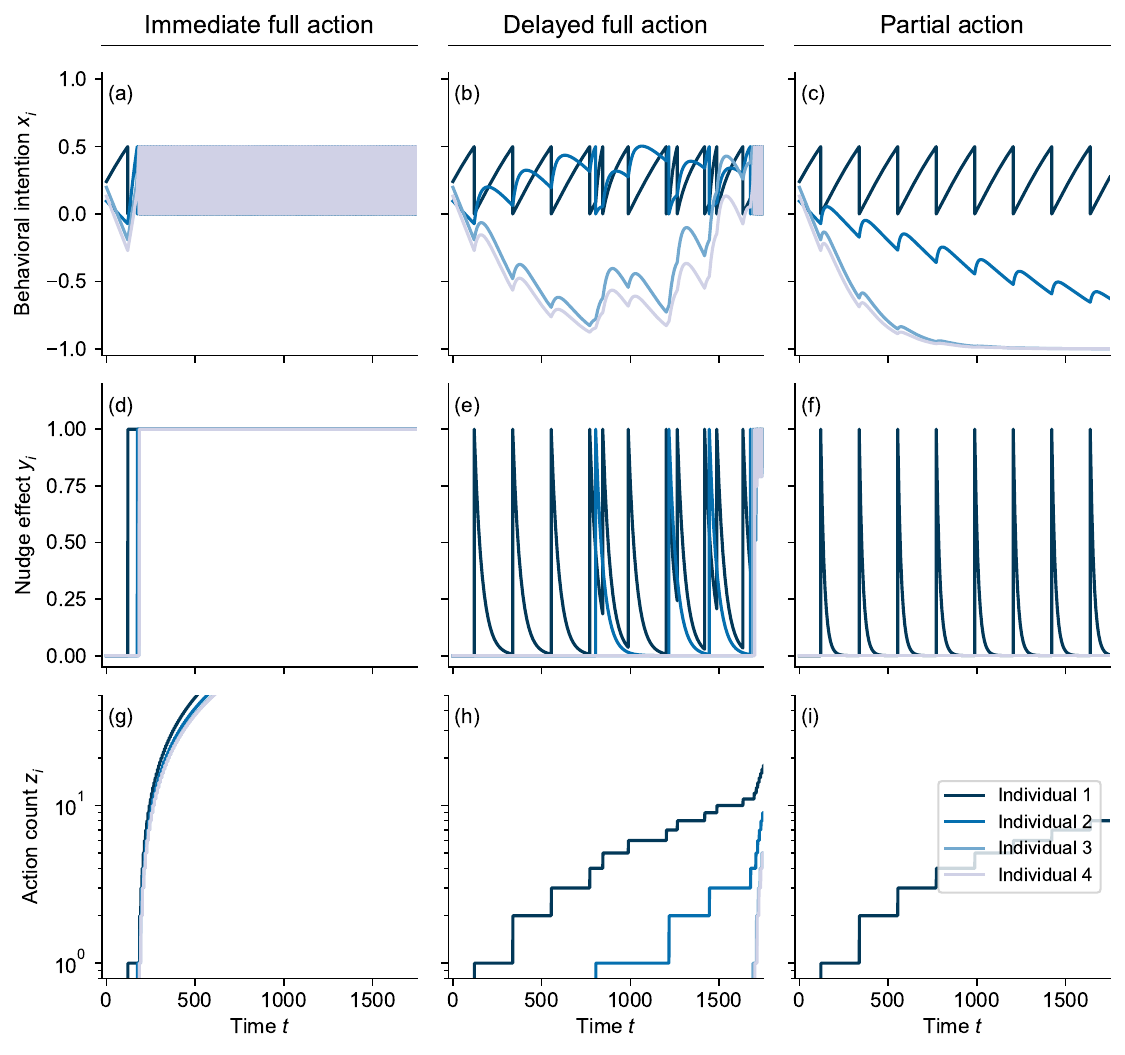}
    \caption{{\bf Effect of immediacy bias on action-taking dynamics.} Behavioral intentions $x_i$ (a--c), nudge effects $y_i$ (d--f), and action counts $z_i$  (g--i) of individuals $i\in\{1,2,3,4\}$, indicated by the colors. The internal attitudes $\alpha_i$ of the four individuals are sampled from $U(-0.4, 1)$. The sample is identical to the values in \autoref{fig:scenarios}c, with $\alpha_1\approx0.60$, $\alpha_2\approx0.44$, $\alpha_3\approx0.37$, and $\alpha_4\approx0.36$.
    Parameters: $\scale_A=1$, $\scale_S=1$, $\scale_C=0.05$, $\cutoff_S=\cutoff_C=0.5$. 
    The initial conditions $x_i(0)\sim U(0, 0.5)$, and $y_i(0)=0$ are identical in all three simulations. We obtained the results for immediate full action, delayed full action, and partial action with immediacy parameters $\immbi=0$, $\immbi=0.023$, and $\immbi=0.05$, respectively. 
    }
    \label{fig:immediacy}
\end{figure}

In the absence of immediacy bias (i.e., $\immbi=0$), the active cluster of a population establishes itself very quickly \autoref{fig:immediacy}a,d,g). When $\immbi=0$, the nudge effects $y_i$ never decay; they follow a step function (\autoref{fig:immediacy}d). Once an individual takes their first action, they exert the strongest possible nudge effect that an individual can exert on the population, and they will continue to do so for all time. Following the first action in a population, all individuals who can be persuaded by a perceived positive social norm to take an action will thus increase their intentions quickly (\autoref{fig:immediacy}a) and join the active cluster in rapid succession (\autoref{fig:immediacy}g).

When $\immbi>0$, nudge effects decay exponentially with time (\autoref{fig:immediacy}e,f), with small (positive) values of $\immbi$ corresponding to a slow decay of nudge effects (\autoref{fig:immediacy}e). Whenever an action occurs, the entire population briefly experiences a strong nudge effect that can lead individuals to increase their intentions (\autoref{fig:immediacy}b). However, since the intensity of the nudge effect decays exponentially, the perceived strong social norm is short-lived. If the perceived social norm does not cause other individuals to take action immediately, it can be necessary for the first action taker to take several actions before a second individual is persuaded to join the active cluster.

When $\immbi$ is sufficiently large, immediacy bias inhibits the emergence of collective action (\autoref{fig:immediacy}c,f,i). The fast decay of nudge effects (\autoref{fig:immediacy}f) prevents the emergence of positive perceived social norms that would lead individuals to increase their intentions (\autoref{fig:immediacy}c) and, eventually, take action. However, individuals who have a very large internal attitude will still take actions not motivated by social influence (\autoref{fig:immediacy}i).

In sum, the number of individuals who take action decreases with the strength of immediacy bias, and full participation in action-taking occurs only when the immediacy parameter is below a certain threshold such that all individuals in the population receive sufficiently strong nudges to act. Above this threshold, the immediacy bias inhibits full action and, at even greater values of $\immbi$, inhibits all action within the collective. We observe these trends with increasing $\immbi$ for almost all combinations of model parameters and initial conditions. The high capacity of the immediacy bias parameter to control the emergence of collective action thus illustrates the crucial role of temporal biases in collective decision-making.

\subsection{\centering D. Modulation of action-taking frequencies}

Our results in \autoref{fig:scenarios} and \autoref{fig:immediacy} highlight the role of intrinsically motivated individuals in kick-starting a behavioral cascade: individuals who will act even in the absence of nudge effects---because their internal attitudes are sufficiently strong---can nudge other less intrinsically motivated individuals to act, even when the latter would otherwise remain inactive. This leads to an observed cascade in which successively activated individuals contribute to an increasing collective nudge. 

\begin{figure}[h!]
    \centering
    \includegraphics[width=0.8\textwidth]{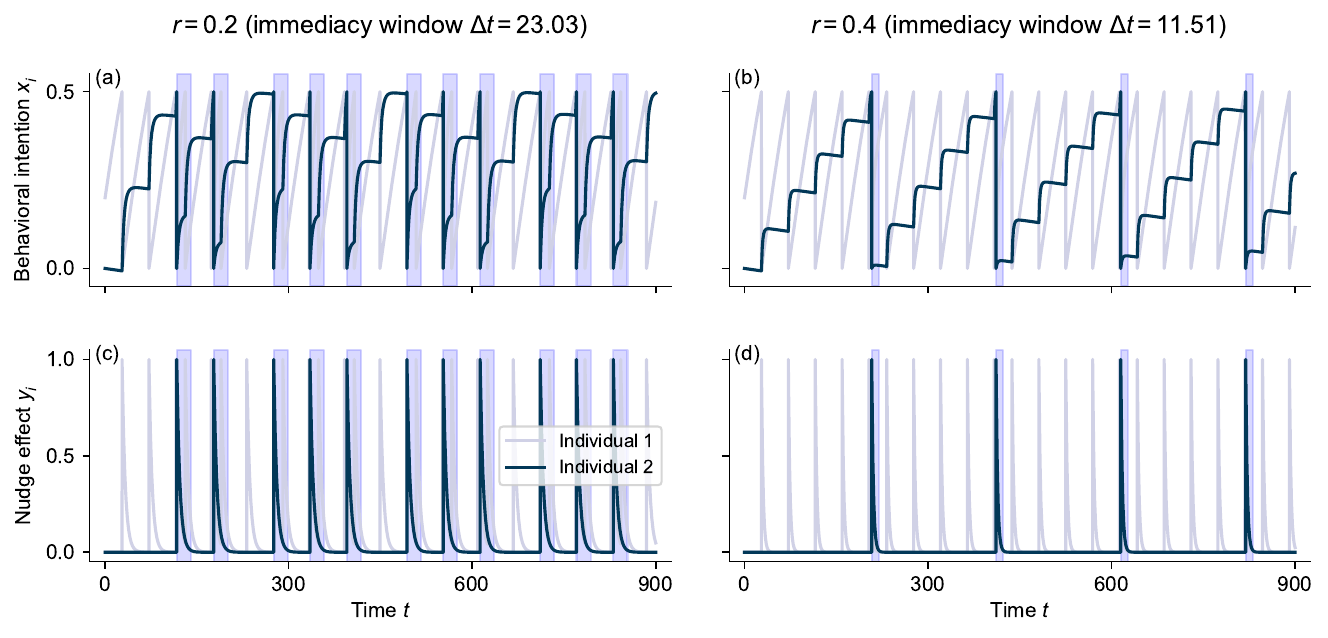}
    \caption{
    \textbf{Impact of the immediacy window on successive action-taking.}
    Behavioral intentions $x_i$ (a,c) and nudge effects $y_i$ (b,d) from simulations with $N=2$ individuals. Colors denote individuals as indicated. Each shaded region denotes the \textit{immediacy window} $\Delta t$ following an action by individual 2, where the immediacy window is the time it takes for an individual $i$'s nudge effect $y_i$ to decay from $1$ to $0.01$ (1\%) in the absence of any additional actions by $i$ or other individuals (i.e., $\Delta t = -\log (0.01)/\immbi$).
    Other parameters: $\cutoff_S=0.5$, $\cutoff_C=0.5$, $\alpha_1=1$, $\alpha_2=0.49$, $y_1(0)=y_2(0)=0$, $x_1(0)=0.2$, $x_2(0)=0$.
    }
    \label{fig:accordion}
\end{figure}

The build-up of the collective nudge is naturally sensitive to the immediacy bias of the population (\autoref{fig:accordion}). To explain how immediacy bias modulates the dynamics of behavioral cascades, we consider a population in the absence of initial nudge effects, and we introduce the notion of an \textit{immediacy window}, which we define as the time it takes for the nudge effect $y_i$ of individual $i$ to decay from 1 to 0.01 (1\%) in the absence of any additional actions by $i$ or other individuals. If immediacy bias is weak (i.e., $\immbi$ is small) and therefore the immediacy window is long, and there is an underlying diversity of intrinsic attitudes among the population, then the feed-forward cascade results in both more effective recruitment to action and an acceleration in action taking (\autoref{fig:accordion}a,c). In contrast, if the immediacy window is short (i.e., $\immbi$ large) and the internal attitudes are appropriately spaced, then the most innately motivated individual will act at approximately the frequency set by their internal motivation alone between the actions of others (\autoref{fig:accordion}b,d). Each action by the most innately motivated individual causes a pulse-like increase in behavioral intentions of other individuals (\autoref{fig:accordion}d), which leads them to also act in an approximately periodic manner (\autoref{fig:accordion}b). If, however, the immediacy window is of intermediate length (i.e., $r$ intermediate), then the timing of individual actions may become less tightly correlated without entering a periodic, or approximately periodic, regime. 
In sum, the interactions among individuals with different internal attitudes, the immediacy window, and the nudge effect may give rise to interesting periodicity in the frequency modulation of action-taking at the population level. 

\section{\centering IV. DISCUSSION}

We have developed a mathematical model to explore the dynamics of collective action in human populations. Our model incorporates psychological nuances grounded in the theory of planned behavior and is informed by well-established temporal biases in human perception. We have used this model to study the social and psychological regimes in which actions propagate within a population. 

We find that the three components of the theory of planned behavior---internal attitudes, perceived social norms, and perceived behavioral control---affect collective dynamics in distinct ways. Varying the strength of social norms or behavioral control results in a gradual increase in the frequency of action-taking; in contrast, varying the strength of internal attitudes results in a phase transition, such that individuals remain inactive until their attitudes cross a critical threshold. The perceptual bias of immediacy also exhibits a threshold effect: while individuals with a negative internal attitude never take action if the immediacy bias is strong, they can be compelled by a perceived social norm to take an action if the immediacy bias is weak enough for the nudge effects to persist for a sufficiently long window of time. Intriguingly, the transient dynamics of behavioral intentions and nudge effects can look similar between scenarios where all individuals eventually take action (`delayed full action' in \autoref{fig:immediacy}) and scenarios where only a subset of individuals take action (`partial action' in \autoref{fig:immediacy}). It can thus be difficult to predict sustained collective action from the initial activity of individuals alone. In this regard, the dynamics of our model mirror those of real-world societal trends and social movements---whose long-term success or failure is often unpredictable in their early stages.

Our model of social behavior exhibits dynamics that resemble those of neuronal networks.
In our model, individuals act when they are nudged by the actions of their neighbors; similarly, in an integrate-and-fire neuronal model, neurons receive nudges from other neurons and activate (i.e., fire) when they reach a threshold \cite{abbott1999lif}. A network of neurons also exhibits patterns of recruitment and oscillations similar to our behavioral model \cite{brunel2000network}. Moreover, neuronal networks can display temporal biases during cognitive tasks---mimicking our model's immediacy effect---driven by plasticity rules that change connectivity strength between neurons \cite{murdock1962,ballintyn2019,greene2000}. These parallels between our model and models of neural networks suggest several directions for future work. For example, while we have assumed that an individual's action always nudges others toward action-taking, nudges could instead increase or decrease the recipients' motivation to act, mimicking the excitatory and inhibitory properties of neurons. Likewise, while we have assumed that an individual can take actions continuously, a neuron cannot fire for a short period following its activation. Analogous constraints may be relevant to human behavior: for example, an individual's perceived behavioral control could decline temporarily after taking an action. Finally, neuronal networks display additional temporal biases, such as primacy biases in which initial inputs are weighed more heavily. In the context of neuronal networks, this can result in dynamic changes in connectivity between neurons that may be relevant to human behavior \cite{fiebig2017,greene2000}.

We have shown that immediacy bias alone can give rise to complex transient dynamics (\autoref{fig:immediacy}) and frequency modulations (\autoref{fig:accordion}). We therefore anticipate that extending our model to include other types of observational biases, such as primacy bias, would result in relevant changes to the collective dynamics that mimic human behavior. Empirical studies in psychology and sociology have shown that behavioral context and experience can shape the relative propensities towards different types of biases\cite{crano1977primacy, miller1959recency, han2021recency}: for example, primacy effects tend to drive investment behaviors \cite{arikan2019primacy}, whereas immediacy effects tend to drive food consumption choices \cite{garbinsky2014interference}; the relative impacts of these biases can depend on the form and strength of social connections \cite{bhargave2015my}. Moreover, the relative importance of different biases may vary over time, even within a narrow question or domain (e.g., investments or food consumption): in the context of consumer choices, for instance, primacy bias tends to be strong when all available options are undesirable but weakens as the options become more attractive overall \cite{li2009best}. Future work could explore how the emergence and dynamics of behavioral cascades in our model are impacted by such context-dependent, temporally varying cognitive biases, as well as interactions between these biases.

\newpage
For simplicity, we have assumed that every individual can observe the actions of all others in the population. While this assumption may be appropriate for small, close-knit communities, it is unlikely to hold in larger, structured societies, where individuals can only sample the actions of their neighbors from time to time. Future work could explore between-individual or temporal variation in the observability, uncertainty, or interpretations of observed actions, perhaps in heterogeneous networks. In addition, we have chosen to model elements of the theory of planned behavior that are thought to directly shape behavioral intentions: internal attitudes, perceived social norms, and perceived behavioral control. However, other, secondary elements of the theory---such as normative beliefs and motivation to comply, which are thought to shape the perceived social norm\cite{ajzen1991theory}---may also be relevant depending on the behavior of interest and domain of application. Understanding the effects of these additional elements remains an important direction for future work.

\vspace{-0.75ex}
\section{ACKNOWLEDGMENTS}
We thank The Banff International Research Station for Mathematical Innovation and Discovery (BIRS) for supporting the Workshop on Building Networks:~Women in Complex and Nonlinear Systems where our collaboration formed and the ideas we have developed were initially proposed. We thank Irina Popovici, Namrata Shukla, Rebecca Hardenbrook, Xie He, Peter Mucha, and Anna Vasenina for fruitful discussions. ACS was supported by funding to Peter J.~Mucha's research group at Dartmouth College, including the Army Research Office MURI award (W911NF-18-1-0244) and startup funds provided by Dartmouth College. MK is supported by a Postdoctoral Fellowship Award from the James S.~McDonnell Foundation (doi:10.37717/2021-3209). TLE is supported by NIH BRAIN Initiative 1K99NS127855-01A1.

\section{APPENDIX}

\subsection{A. Lower bound on the critical internal attitude}

In the absence of initial nudge effects (i.e., $y_i(0)=0$ for all individuals $i$), we have $\beta_i=0$ at $t=0$ for all individuals $i$. The time evolution of internal attitudes then follows
\begin{equation+} 
    \dot x_{i} = \left[\scale_A\alpha_i  - \scale_S\cutoff_S\right] \scale_C\cutoff_C g(x_{i,j})\;, \qquad i\in\{1,\dots,N\} \;. \nonumber
\end{equation+} 
Assuming positive scale parameters $\scale_A$, $\scale_S$, and $\scale_C$ and shift parameters $\cutoff_S$ and $\cutoff_C$, the growth of the internal attitude $\dot x_{i}$ prior to any action being taken is negative or 0 if 
\begin{equation+} 
    \scale_S\cutoff_S\geq\scale_A\alpha_i \;.  \nonumber
\end{equation+} 
If $\scale_A\leq \chi_A=\scale_S\cutoff_S/\max\{\alpha_i\}$, the internal attitudes of all individuals are decreasing and all individuals remain inactive for all time.

\small
\bibliographystyle{unsrt}
\bibliography{main}

\end{document}